# Partially Ordered Two-way Büchi Automata[*]

Manfred Kufleitner     Alexander Lauser

FMI, Universität Stuttgart, Germany
`{kufleitner,lauser}@fmi.uni-stuttgart.de`

**Abstract.** We introduce partially ordered two-way Büchi automata and characterize their expressive power in terms of fragments of first-order logic FO[<]. Partially ordered two-way Büchi automata are Büchi automata which can change the direction in which the input is processed with the constraint that whenever a state is left, it is never re-entered again. Nondeterministic partially ordered two-way Büchi automata coincide with the first-order fragment $\Sigma_2$. Our main contribution is that deterministic partially ordered two-way Büchi automata are expressively complete for the first-order fragment $\Delta_2$. As an intermediate step, we show that deterministic partially ordered two-way Büchi automata are effectively closed under Boolean operations.

A small model property yields coNP-completeness of the emptiness problem and the inclusion problem for deterministic partially ordered two-way Büchi automata.

## 1 Introduction

The original motivation of Büchi automata was to decide monadic second order logic over infinite words [3, 19]. A Büchi automaton is a nondeterministic finite automaton which accepts an infinite word if there is a run of the automaton on the word such that some final state appears infinitely often. It is well-known that deterministic Büchi automata are less expressive than nondeterministic ones. Today, Büchi automata have become one of the most important tools for formal verification of sequential finite state systems, see e.g. [1, 4]. A generalization are two-way Büchi automata which have the same expressive power as one-way Büchi automata [13], but the conversion of a two-way Büchi automaton into a one-way Büchi automaton may require an exponential blow-up [8].

[*]This work was supported by the German Research Foundation (DFG), grant DI 435/5-1.

An automaton is *partially ordered* if there exists a partial ordering of the states which respects the transition relation, i.e., in any computation once a state is left, it is never re-entered again. Partially ordered automata are also known as 1-weak, very weak or linear automata, cf. [6]. The last name comes from the fact that we may also use a linear ordering of the states instead of a partial ordering. The drawback of using linear orders is that this distorts the maximal length of chains of states, which in some situations is a significant parameter. Partially ordered one-way Büchi automata can be used for a characterization of the common fragment of the two temporal logics ACTL and LTL [2, 12]. As we will see in Section 3, nondeterministic partially ordered one-way Büchi automata and nondeterministic partially ordered two-way Büchi automata have the same expressive power.

Schwentick, Thérien, and Vollmer introduced partially ordered two-way automata over finite words [15]. They showed that the nondeterministic variant is expressively complete for the fragment $\Sigma_2$ of first-order logic FO = FO[<]; they also showed that deterministic partially ordered two-way automata coincide with the first-order fragment $\Delta_2$. Here, $\Sigma_2$ consists of all FO formulas in prenex normal form with two blocks of quantifiers, starting with existential quantifiers. The fragment $\Delta_2$ comprises all languages $L$ such that both $L$ and its complement are $\Sigma_2$-definable. The class of languages definable in $\Delta_2$ is the largest subclass of $\Sigma_2$ which is closed under complementation. Various characterizations are known for $\Sigma_2$ and $\Delta_2$ over infinite words, see e.g. [5, 7, 18]. When interpreted over infinite word models, the fragment $\Delta_2$ is, in some sense, weaker than over finite words. Over finite words it coincides with the fragment $FO^2$ of first-order logic with only two variables [17] and the languages definable in $\Delta_2$ are precisely so-called unambiguous polynomials [14]. Over infinite words however, $\Delta_2$ is a strict subclass of $FO^2$ and only restricted unambiguous polynomials are definable in $\Delta_2$. Partially ordered two-way automata over finite words have also been characterized by an interval temporal logic [10].

In this paper, we give an extension of the results of Schwentick et al. [15] to infinite words. We introduce *partially ordered two-way (po2) Büchi automata* and we give completeness results for their expressive power:

- $L \subseteq \Gamma^\omega$ is recognized by some nondeterministic partially ordered two-way Büchi automaton if and only if $L$ is definable in $\Sigma_2$ (Theorem 2).

- $L \subseteq \Gamma^\omega$ is recognized by some deterministic partially ordered two-way Büchi automaton if and only if $L$ is definable in $\Delta_2$ (Theorem 13).

An immediate corollary is that deterministic po2-Büchi automata are less expressive than their nondeterministic counterparts. Moreover, it will turn out that nondeterministic partially ordered one-way and nondeterministic po2-Büchi automata have the same expressive power (Theorem 2), whereas deterministic partially ordered one-way Büchi automata are strictly less expressive than deterministic po2-Büchi automata (Example 15). It is decidable whether a regular $\omega$-language is definable in $\Sigma_2$ or in $\Delta_2$, respectively [2]. Hence, it is decidable whether a regular $\omega$-language can be recognized by some nondeterministic po2-Büchi automaton or by some deterministic po2-Büchi automaton, respectively.

Theorem 2 is a straightforward extension of the finite case, whereas the situation for deterministic po2-Büchi automata in Theorem 13 is more involved. The key ingredient is to show that deterministic po2-Büchi automata are closed under Boolean operations (Corollary 10). This is nontrivial since, contrary to the finite case, we cannot run po2-Büchi automata one after another because the computations are infinite. To overcome this problem, we simulate two deterministic po2-Büchi automata in parallel by a product automaton construction. The computations of deterministic po2-Büchi automata exhibit some combinatorial property which allows us to keep track of the head positions of the automata when they disagree on the direction of the head movement.

Finally we give complexity results for some decision problems concerning po2-Büchi automata in Section 5. We show that the inclusion problem as well as the emptiness problem, the universality problem and the equivalence problem for deterministic po2-Büchi automata are coNP-complete. Moreover, we show that nondeterministic po2-Büchi automata have a *small model property* leading to a coNP-completeness result for the emptiness problem for nondeterministic po2-Büchi automata. Similar complexity results for po2-automata over finite words have been shown by Lodaya, Pandya, and Shah [11]. For general one-way Büchi automata (that need not be partially ordered), the inclusion problem and the equivalence problem are PSPACE-complete [16].

The results of this paper were presented at the 15th International Conference on Implementation and Application of Automata (CIAA 2010) in Winnipeg, Canada [9].

## 2 Preliminaries

Throughout, $\Gamma$ denotes a finite alphabet. The set of finite words over the alphabet $A \subseteq \Gamma$ is $A^*$, the set of infinite words over $A$ is $A^\omega$, and $A^\infty = A^* \cup A^\omega$ is the set of finite and infinite words. The empty word is denoted by $\varepsilon$. The *length* of a word $\alpha$ is $|\alpha| \in \mathbb{N} \cup \{\infty\}$ and $\alpha(i)$ is the $i$-th letter of $\alpha$. We have $\alpha = \alpha(1) \cdots \alpha(n)$ if $|\alpha| = n \in \mathbb{N}$ and $\alpha = \alpha(1)\alpha(2) \cdots$ if $|\alpha| = \infty$. The *alphabet* of $\alpha$ is $\text{alph}(\alpha)$. It is the set $\{a \in \Gamma \mid \alpha = ua\beta \text{ for some } u, \beta\}$ of letters occurring in $\alpha$. A position $i$ of $\alpha$ is an *a-position* if $\alpha(i) = a$. For $u, v \in \Gamma^*$ we write $u \leq_s v$ if $u$ is a suffix of $v$, i.e., $v = wu$ for some $w \in \Gamma^*$. A word $v = a_1 \cdots a_n \in \Gamma^*$ is a *scattered subword* of $\alpha \in \Gamma^\infty$, denoted by $v \preccurlyeq \alpha$ if $\alpha = u_1 a_1 \cdots u_n a_n \beta$ for some $u_i \in \Gamma^*$ and $\beta \in \Gamma^\infty$. *Languages* are subsets of $\Gamma^\infty$. In order to emphasize that a language $L \subseteq \Gamma^\omega$ contains only infinite words, we also say that $L$ is an $\omega$-*language*. A *monomial* of degree $k$ is a language of the form $A_1^* a_1 \cdots A_k^* a_k A_{k+1}^*$. It is *unambiguous* if every word $w$ has at most one factorization $w = u_1 a_1 \cdots u_k a_k u_{k+1}$ with $u_i \in A_i^*$. Similarly, an $\omega$-*monomial* of degree $k$ is an $\omega$-language of the form $A_1^* a_1 \cdots A_k^* a_k A_{k+1}^\omega$ and it is *unambiguous* if every word $\alpha$ has at most one factorization $u_1 a_1 \cdots u_k a_k \beta$ with $u_i \in A_i^*$ and $\beta \in A_{k+1}^\omega$. It is a *restricted unambiguous $\omega$-monomial* if $\{a_i, \ldots, a_k\} \nsubseteq A_i$ for all $1 \leq i \leq k$. An ($\omega$-)*polynomial* is a finite union of ($\omega$-)monomials.

## 2.1 Partially ordered two-way Büchi automata

In the following, we give the Büchi automaton counterpart of a two-way automaton. A *two-way Büchi automaton* $\mathcal{A} = (Z, \Gamma, \delta, Z_0, F)$ is given by:

- a finite set of states $Z = X \,\dot\cup\, Y$,
- a finite input alphabet $\Gamma$; the tape alphabet is $\Gamma \,\dot\cup\, \{\triangleright\}$, where the left end marker $\triangleright$ is a new letter,
- a transition relation $\delta \subseteq (Z \times \Gamma \times Z) \cup (Y \times \{\triangleright\} \times X)$,
- a set of initial states $Z_0 \subseteq Z$, and
- a set of final states $F \subseteq Z$.

The states $Z$ are partitioned into "ne$X$t-states" $X$ and "$Y$esterday-states" $Y$. The idea is that states in $X$ are entered with a right-move of the head while states in $Y$ are entered with a left-move. For $(z, a, z') \in \delta$ we frequently use the notation $z \xrightarrow{a} z'$. On input $\alpha = a_1 a_2 \cdots \in \Gamma^\omega$ the tape is labeled by $\triangleright \alpha$, i.e., positions $i \geq 1$ are labeled by $a_i$ and position 0 is labeled by $\triangleright$. A *configuration* of the automaton is a pair $(z, i)$ where $z \in Z$ is a state and $i \in \mathbb{N}$ is the current position of the head. A *transition* $(z, i) \vdash_\mathcal{A} (z', j)$ on configurations $(z, i)$ and $(z', j)$ exists if

- $z \xrightarrow{a} z'$ for some $a \in \Gamma \cup \{\triangleright\}$ such that $i$ is an $a$-position, and
- $j = i + 1$ if $z' \in X$, and $j = i - 1$ if $z' \in Y$.

The $\triangleright$-position can only be encountered in a state from $Y$ and left via a state from $X$. In particular, $\mathcal{A}$ can never overrun the left end marker $\triangleright$. Due to the partition of the states $Z$, we can never have a change in direction without changing the state. A *computation* of $\mathcal{A}$ on input $\alpha$ is an infinite sequence of transitions

$$(z_1, i_1) \vdash_\mathcal{A} (z_2, i_2) \vdash_\mathcal{A} (z_3, i_3) \vdash_\mathcal{A} \cdots$$

such that $z_1 \in Z_0$ and $i_1 = 1$. It is *accepting* if there exists some state $y \in F$ which occurs infinitely often in this computation. A word $\alpha$ is *accepted* by $\mathcal{A}$ if there is an accepting computation of $\mathcal{A}$ on input $\alpha$. As usual, the language recognized by $\mathcal{A}$ is $L(\mathcal{A}) = \{\alpha \in \Gamma^\omega \mid \mathcal{A} \text{ accepts } \alpha\}$.

A two-way Büchi automaton is *deterministic* if $|Z_0| = 1$ and if for every state $z \in Z$ and every letter $a \in \Gamma \cup \{\triangleright\}$ there is at most one $z' \in Z$ with $z \xrightarrow{a} z'$. A two-way Büchi automaton is *complete* if for every state $z \in Z$ and every letter $a \in \Gamma$ there is at least one $z' \in Z$ with $z \xrightarrow{a} z'$, and for every $z \in Y$ there is at least one $z' \in X$ with $z \xrightarrow{\triangleright} z'$.

We are now ready to define *partially ordered* two-way Büchi automata. We use the abbreviation "po2" for "partially ordered two-way". The set of states $Z$ in a *po2-Büchi automaton* is equipped with a partial order $\sqsubseteq$ such that $z \xrightarrow{a} z'$ implies $z \sqsubseteq z'$ for all $z, z' \in Z$ and $a \in \Gamma$. In po2-Büchi automata, every computation enters a state at most once and it defines a finite chain of states. Thus every computation has a unique state $z \in Z$ which occurs infinitely often and this state is maximal among all states in the computation. Moreover, $z \in X$ since the automaton cannot loop in a left-moving state forever. We call this state $z$ *stationary*. A computation is accepting if and only if its stationary state $z$ is a final state. Note that every partially ordered two-way Büchi automaton can be converted into a complete partially ordered two-way Büchi automaton recognizing the same language by adding a new sink state and redirecting missing transitions to the sink state.

**Example 1** Consider the following two-way Büchi automaton $\mathcal{A}$ with $X$-states $\{x_0, x_1\}$ and $Y$-states $\{y_1, y_2, y_3\}$ over the alphabet $\Gamma = \{a, b, c, d\}$:

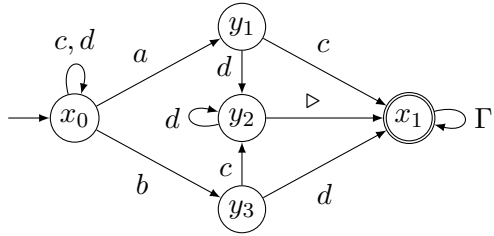

This automaton is deterministic and partially ordered. The language accepted is $L(\mathcal{A}) = \{c, d\}^* \{ca, db\} \Gamma^* \cup \{d\}^* \{da, cb\} \Gamma^*$. $\diamond$

## 2.2 Fragments of first-order logic

In first-order logic FO[<] = FO, words are interpreted as labeled linear orders and variables range over positions of the word. Atomic formulas of FO are $\top$ for true, $\lambda(x) = a$ and $x < y$ for variables $x, y$ and letters $a \in \Gamma$. A modality $\lambda(x) = a$ means that $x$ is an $a$-position and $x < y$ says that the position $x$ is smaller than $y$. Formulas can be composed by Boolean connectives, existential quantification $\exists x \colon \varphi$, and universal quantification $\forall x \colon \varphi$ for a formula $\varphi \in \text{FO}$. The semantics of these operators are as usual. For every formula there exists an equivalent formula in prenex normal form by renaming variables and moving quantifiers to the front. The fragment $\Sigma_2$ (resp. $\Pi_2$) consists of all formulas in FO which are in prenex normal form with one quantifier alternation, starting with a block of existential (resp. universal) quantifiers. We identify equivalent formulas and hence it makes sense to define $\Delta_2 = \Sigma_2 \cap \Pi_2$, which contains all formulas in $\Sigma_2$ which have an equivalent formula in $\Pi_2$. Note that the notion of equivalence of formulas depends on the models we allow. More concretely, the fragment $\Delta_2$ depends on whether we interpret formulas over $\Gamma^*$ or over $\Gamma^\omega$, cf. [7, 17]. Unless stated otherwise, we consider infinite word models in the remainder of the paper. A *sentence* is a formula in FO without free variables. For every word $\alpha$, the truth value of a sentence $\varphi$ is well-defined and we write $\alpha \models \varphi$ if $\varphi$ is true when interpreted over $\alpha$. For a sentence $\varphi$, the $\omega$-*language defined by* $\varphi$ is $L(\varphi) = \{\alpha \in \Gamma^\omega \mid \alpha \models \varphi\}$. A language $L \subseteq \Gamma^\omega$ is *definable in a fragment* $\mathcal{F}$ if there exists a sentence $\varphi \in \mathcal{F}$ such that $L(\varphi) = L$.

## 3 Nondeterministic po2-Büchi Automata

This section contains the characterization of nondeterministic po2-Büchi automata by the first-order fragment $\Sigma_2$ over infinite words. As a byproduct, we will see that nondeterministic partially ordered one-way Büchi automata (i.e., $Y = \emptyset$) have the same expressive power as nondeterministic po2-Büchi automata. The proof of this result is a straightforward extension of the finite case [15]. It is presented here for the sake of completeness.

**Theorem 2** *Let $L \subseteq \Gamma^\omega$. The following assertions are equivalent:*

1. *$L$ is recognized by a nondeterministic po2-Büchi automaton.*
2. *$L$ is definable in $\Sigma_2$.*
3. *$L$ is recognized by a nondeterministic partially ordered Büchi automaton.*

*Proof:* "1 $\Rightarrow$ 2": Let $L$ be recognized by a po2-Büchi automaton. From Lemma 3 below (with $\mathcal{A} = \mathcal{B}$) we get that $L$ is an $\omega$-polynomial. The claim follows since $\omega$-polynomials are $\Sigma_2$-definable.

"2 $\Rightarrow$ 3": Let $L$ be definable in $\Sigma_2$. Then $L$ is an $\omega$-polynomial [18]. Now, the $\omega$-monomial $A_1^* a_1 \cdots A_k^* a_k A_{k+1}^\omega$ is recognized by the following Büchi automaton:

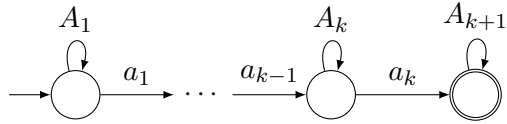

Every $\omega$-polynomial is recognized by a finite union of such automata.

The implication "3 $\Rightarrow$ 1" is trivial. $\square$

For the proof of Theorem 2, a slightly weaker property than the one given in Lemma 3 would suffice. We prove a more general statement for later use in Section 5.

**Lemma 3** *Let $\mathcal{A}$ and $\mathcal{B}$ be complete po2-Büchi automata and let $n_\mathcal{A}$ and $n_\mathcal{B}$ be the lengths of the longest chains of states of $\mathcal{A}$ and $\mathcal{B}$, respectively. For every $\alpha \in L(\mathcal{A}) \cap L(\mathcal{B})$ there exists an $\omega$-monomial $P_\alpha$ of degree at most $n_\mathcal{A} + n_\mathcal{B} - 2$ such that $\alpha \in P_\alpha \subseteq L(\mathcal{A}) \cap L(\mathcal{B})$. In particular, $L(\mathcal{A}) \cap L(\mathcal{B})$ is an $\omega$-polynomial.*

*Proof:* Let $\alpha \in L(\mathcal{A}) \cap L(\mathcal{B})$ and consider an accepting computation of $\mathcal{A}$ and an accepting computation of $\mathcal{B}$. For these computations, we define the factorization $\alpha = u_1 a_1 \cdots u_k a_k \beta$ with $a_i \in \Gamma$, $u_i \in \Gamma^*$, and $\beta \in \Gamma^\omega$ such that the positions of the markers $a_i$ are exactly those where a state change happens in at least one of the computations. In each traversal of one of the factors $u_i$ and $\beta$, the letters in these factors correspond to self-loops on the respective states in both computations. Hence, $P_\alpha = A_1^* a_1 \cdots A_k^* a_k B^\omega$ with $A_i = \mathrm{alph}(u_i)$ and $B = \mathrm{alph}(\beta)$ satisfies $\alpha \in P_\alpha$, $P_\alpha \subseteq L(\mathcal{A})$, and $P_\alpha \subseteq L(\mathcal{B})$. Therefore, $L(\mathcal{A}) \cap L(\mathcal{B}) = \bigcup \{P_\alpha \mid \alpha \in L(\mathcal{A}) \cap L(\mathcal{B})\}$ is an $\omega$-polynomial since there are only finitely many $\omega$-monomials of degree at most $n_\mathcal{A} + n_\mathcal{B} - 2$. $\square$

**Corollary 4** *For a given regular $\omega$-language $L$ it is decidable whether $L$ can be recognized by some nondeterministic po2-Büchi automaton (or equivalently, by some nondeterministic partially ordered Büchi automaton).*

*Proof:* By Theorem 2, this decision problem is equivalent to $L$ being $\Sigma_2$-definable. The latter is known to be decidable [2]. □

## 4 Deterministic po2-Büchi Automata

In this section we prove our main result: Deterministic po2-Büchi automata are expressively complete for the first-order fragment $\Delta_2$ over infinite words. The proof is based on a characterization of $\Delta_2$ in terms of restricted unambiguous $\omega$-monomials [7]. The key ingredient is to show that deterministic po2-Büchi automata are effectively closed under Boolean operations (Corollary 10). We first show the closure under complementation in Lemma 5. At first sight, this result is surprising since deterministic two-way Büchi automata are not closed under complementation. In Proposition 9 we show that deterministic po2-Büchi automata are closed under union and intersection. The idea is to simulate two automata simultaneously. The main problem is a dissent on the direction of the head movement. This problem is solved using a combinatorial property of the computation of a deterministic po2-Büchi automata which is formulated in Proposition 7 and translated into an automaton construction in Lemma 8.

**Lemma 5** *If $L$ is recognized by a complete deterministic po2-Büchi automaton $\mathcal{A} = (Z, \Gamma, \delta, z_0, F)$, then $\Gamma^\omega \setminus L$ is recognized by the complete deterministic po2-Büchi automaton $\overline{\mathcal{A}} = (Z, \Gamma, \delta, z_0, Z \setminus F)$.*

*Proof:* Let $\mathcal{A}$ be a complete deterministic po2-Büchi automaton recognizing $L$. For every word $\alpha$ there is a unique computation of $\mathcal{A}$. Therefore, every word $\alpha$ uniquely determines a stationary state $x_\alpha$ and $\alpha$ is accepted if and only if $x_\alpha$ is final. Thus, complementing the set of final states yields a deterministic po2-Büchi automaton $\overline{\mathcal{A}}$ for the complement of $L(\mathcal{A})$. □

We now turn to the closure under union and intersection. Prior to this, we need some more notation which we will use later for a product automaton construction.

Let $\alpha \in \Gamma^\infty$ be a finite or infinite word. For a scattered subword $v \preccurlyeq \alpha$ with $v = a_1 \cdots a_m \in \Gamma^+$, the *$v$-prefix factorization* of $\alpha$ is $\alpha = u_1 a_1 \cdots u_m a_m \beta$ with $u_i \in (\Gamma \setminus \{a_i\})^*$ and $\beta \in \Gamma^\infty$. For every $v \preccurlyeq \alpha$, the $v$-prefix factorization of $\alpha$ is unique. The next lemma justifies the name.

**Lemma 6** *Let $\alpha \in \Gamma^\infty$, let $v = a_1 \cdots a_m \in \Gamma^+$, and let $\alpha = u_1 a_1 \cdots u_m a_m \beta$ be the $v$-prefix factorization. Then for all $k \in \{1, \ldots, n\}$, the word $u_1 a_1 \cdots u_k a_k$ is the shortest prefix of $\alpha$ such that $a_1 \cdots a_k \preccurlyeq u_1 a_1 \cdots u_k a_k$.*

*Proof:* We have $a_1 \notin \mathrm{alph}(u_1)$. Hence, the claim is true for $k = 1$. Let now $k > 1$. Obviously, $a_1 \cdots a_k \preccurlyeq u_1 a_1 \cdots u_k a_k$. Assume $a_1 \cdots a_k \preccurlyeq u_1 a_1 \cdots u_{k-1} a_{k-1} u_k$. Since $a_k \notin \mathrm{alph}(u_k)$, we conclude $a_1 \cdots a_k \preccurlyeq u_1 a_1 \cdots u_{k-1} a_{k-1}$ and hence, $a_1 \cdots a_{k-1} \preccurlyeq u_1 a_1 \cdots u_{k-1}$. This contradicts the induction hypothesis. Therefore, $u_1 a_1 \cdots u_k a_k$ is the shortest prefix of $\alpha$ such that $a_1 \cdots a_k \preccurlyeq u_1 a_1 \cdots u_k a_k$. □

Consider the $v$-prefix factorization $\alpha = u_1 a_1 \cdots u_m a_m \beta$ for $v = a_1 \cdots a_m \in \Gamma^+$. A word $w \in \Gamma^*$ is $k$-*prefix compatible* for $k \in \{1, \ldots, m\}$, if

$$a_k \cdots a_m \preccurlyeq w \leq_s u_k a_k \cdots u_m a_m.$$

One can think of the scattered subword property $a_k \cdots a_m \preccurlyeq w$ as a lower bound for $|w|$, and $w$ being a suffix of $u_k a_k \cdots u_m a_m$ yields an upper bound for $|w|$, i.e., if $w$ is $k$-prefix compatible, then, in some sense, $w$ is neither too short nor too long. As shown in the following proposition, it is easily possible to keep track of $k$-prefix compatible suffixes.

**Proposition 7** *Let $\alpha \in \Gamma^\infty$, let $v = a_1 \cdots a_m \in \Gamma^+$ and let $\alpha = u_1 a_1 \cdots u_m a_m \beta$ be the $v$-prefix factorization, i.e., $a_i \notin \mathrm{alph}(u_i)$. Let $w \in \Gamma^*$ and $c \in \Gamma$ be such that $cw \leq_s u_1 a_1 \cdots u_m a_m$. Suppose $1 < k \leq m$ and $1 \leq \ell < m$ for $k, \ell \in \mathbb{N}$.*

1. *If $w$ is $1$-prefix compatible, then $cw$ is $1$-prefix compatible.*
2. *If $w$ is $k$-prefix compatible and $c = a_{k-1}$, then $cw$ is $(k-1)$-prefix compatible.*
3. *If $w$ is $k$-prefix compatible and $c \neq a_{k-1}$, then $cw$ is $k$-prefix compatible.*
4. *If $cw$ is $\ell$-prefix compatible and $c = a_\ell$, then $w$ is $(\ell+1)$-prefix compatible.*
5. *If $cw$ is $\ell$-prefix compatible and $c \neq a_\ell$, then $w$ is $\ell$-prefix compatible.*
6. *If $cw$ is $m$-prefix compatible and $c = a_m$, then $w = \varepsilon$.*
7. *If $cw$ is $m$-prefix compatible and $c \neq a_m$, then $w$ is $m$-prefix compatible.*

*Proof:* "1": If $a_1 \cdots a_m \preccurlyeq w$, then $a_1 \cdots a_m \preccurlyeq cw$. By definition, $cw$ is a suffix of $u_1 a_1 \cdots u_m a_m$. Hence, $cw$ is 1-prefix compatible.

"2": If $a_k \cdots a_m \preccurlyeq w$, then $a_{k-1} a_k \cdots a_m \preccurlyeq a_{k-1} w$. With $w \leq_s u_k a_k \cdots u_m a_m$ and $a_{k-1} w \leq_s u_1 a_1 \cdots u_m a_m$ we get $a_{k-1} w \leq_s a_{k-1} u_k a_k \cdots u_m a_m \leq_s u_{k-1} a_{k-1} \cdots u_m a_m$. Therefore, $a_{k-1} w$ is $(k-1)$-prefix compatible.

"3": If $a_k \cdots a_m \preccurlyeq w$, then $a_k \cdots a_m \preccurlyeq cw$. Moreover, if $w \leq_s u_k a_k \cdots u_m a_m$ and $cw \leq_s u_1 a_1 \cdots u_m a_m$, then $cw \leq_s a_{k-1} u_k a_k \cdots u_m a_m$. With $c \neq a_{k-1}$ we conclude $cw \leq_s u_k a_k \cdots u_m a_m$. This shows that $cw$ is $k$-prefix compatible.

"4": If $a_\ell \cdots a_m \preccurlyeq cw$, then $a_{\ell+1} \cdots a_m \preccurlyeq w$. From $a_\ell w \leq_s u_\ell a_\ell \cdots u_m a_m$ we get $a_\ell w \leq_s a_\ell u_{\ell+1} a_{\ell+1} \cdots u_m a_m$ because $a_\ell \notin \mathrm{alph}(u_\ell)$. Thus $w \leq_s u_{\ell+1} a_{\ell+1} \cdots u_m a_m$ which yields that $w$ is $(\ell+1)$-prefix compatible.

"5": If $a_\ell \cdots a_m \preccurlyeq cw$ and $c \neq a_\ell$, then $a_\ell \cdots a_m \preccurlyeq w$. If $cw \leq_s u_\ell a_\ell \cdots u_m a_m$, then $w \leq_s u_\ell a_\ell \cdots u_m a_m$. Hence, $w$ is $\ell$-prefix compatible.

"6": We have $a_m w \leq_s a_m$ since $a_m w \leq_s u_m a_m$ and $a_m \notin \mathrm{alph}(u_m)$. Thus $w = \varepsilon$.

"7": If $a_m \preccurlyeq cw$ and $c \neq a_m$, then $a_m \preccurlyeq w$; and if $cw \leq_s u_m a_m$, then $w \leq_s u_m a_m$. Therefore, $w$ is $m$-prefix compatible. □

Next, we give an "automaton version" of Lemma 7. Let $u_1 a_1 \cdots u_m a_m$ be a $v$-prefix factorization for $v = a_1 \cdots a_m$. Suppose a deterministic po2-Büchi automaton $\mathcal{A}$ starts some computation on $u_1 a_1 \cdots u_m a_m \beta$ at the position $i_0 = |u_1 a_1 \cdots u_m a_m|$ with a left-move of the head. We construct an equivalent deterministic po2-Büchi automaton $\mathcal{C}$ which in some sense is aware of the first time the position $i_0$ is re-visited.

**Lemma 8** *Let $\mathcal{A}$ be a deterministic po2-Büchi automaton with states $Z = X \dot\cup Y$. For every $v = a_1 \cdots a_m \in \Gamma^+$ there effectively exists a deterministic po2-Büchi automaton $\mathcal{C}$ with state set $Z_{\mathcal{C}} = Z \times \{v\} \times \{1, \ldots, m\}$ such that for all $\alpha \in \Gamma^\omega$ with a $v$-prefix factorization $\alpha = u_1 a_1 \cdots u_m a_m \beta$ the following property holds: If*

$$(z_0, i_0) \vdash_{\mathcal{A}} (z_1, i_1) \vdash_{\mathcal{A}} \cdots \vdash_{\mathcal{A}} (z_n, i_n)$$

*is a sequence of transitions of $\mathcal{A}$ for some $n \geq 1$ satisfying $i_0 = i_n = |u_1 a_1 \cdots u_m a_m|$ and $i_t < i_n$ for all $1 \leq t < n$, then*

$$\bigl((z_1, v, k_1), i_1\bigr) \vdash_{\mathcal{C}} \cdots \vdash_{\mathcal{C}} \bigl((z_n, v, k_n), i_n\bigr)$$

*is a sequence of transitions of $\mathcal{C}$ with $k_1 = k_n = m$ such that there exists no $1 \leq t < n$ with $z_t \in X$, $k_t = m$, and $\alpha(i_t) = a_m$.*

*Proof:* A position $i \leq i_n$ is $k$-*prefix compatible* if the factor of $\alpha$ induced by the interval of positions $\{j \in \mathbb{N} \mid i \leq j \leq i_n\}$ is $k$-prefix compatible. For $z \in Z$, we define $\xi(z) = 0$ if $z \in X$, and $\xi(z) = 1$ if $z \in Y$. For every transition $z \xrightarrow{c} z'$ in $\mathcal{A}$ and every $k \in \{1, \ldots, m\}$, we give a $k' \in \{1, \ldots, m\}$ such that $(z, v, k) \xrightarrow{c} (z', v, k')$ in $\mathcal{C}$. Moreover, every transition $\bigl((z, v, k), i\bigr) \vdash_{\mathcal{C}} \bigl((z', v, k'), j\bigr)$ will satisfy the following invariant:

If $i + \xi(z)$ is $k$-prefix compatible, then $j + \xi(z')$ is $k'$-prefix compatible.

The claim of the lemma then follows by Proposition 7 (6) since $z_1 \in Y$ and $i_1 + \xi(z_1) = i_0$ is $m$-prefix compatible. In particular, we do not have to treat the case $z \in X$, $k = m$, and $c = a_m$.

We now construct $k'$. For this purpose we first define the prefix compatibility $\ell$ of the head position before the transition is made. If $z \in X$, then we set $\ell = k$ by definition of $\xi$. If $z \in Y$ and $c = a_{k-1}$, then $\ell = k - 1$ by Proposition 7 (2). In the remaining case $z \in Y$ and $c \neq a_{k-1}$, we set $\ell = k$ by Proposition 7 (1) and Proposition 7 (3). Finally, we define $k'$. If $z' \in Y$, then $k' = \ell$ by definition of $\xi$. If $z' \in X$ and $c = a_k$, then $k' = \ell + 1$ by Proposition 7 (4); otherwise we have $z' \in X$ and $c \neq a_k$, and we set $k' = \ell$ by Proposition 7 (5) and Proposition 7 (7).

Note that once the counter value changes, there must be a change of direction (and hence a change of state in $\mathcal{A}$) before we may encounter the same value again. Therefore $\mathcal{C}$ is partially ordered. □

**Proposition 9** *Let $\mathcal{A}_1$ and $\mathcal{A}_2$ be complete deterministic po2-Büchi automata. There effectively exists a deterministic po2-Büchi automaton $\mathcal{B}$ such that $L(\mathcal{B}) = L(\mathcal{A}_1) \cup L(\mathcal{A}_2)$ (resp. $L(\mathcal{B}) = L(\mathcal{A}_1) \cap L(\mathcal{A}_2)$).*

*Moreover, let $n_i$ be the number of states in $\mathcal{A}_i$, let $m_i$ be the length of a maximal chain of next-states in $\mathcal{A}_i$, and let $m = m_1 + m_2 - 2$. Then $\mathcal{B}$ has at most $3m n_1 n_2 |\Gamma|^{m+1}$ states.*

*Proof:* We start with some intuition on our construction before we give the actual proof. The idea is that $\mathcal{B}$ simulates both automata simultaneously by a product automaton construction in what we call the *synchronous mode*. However, we have to handle the case when the automata disagree on the direction in which the input is processed. In this case we switch to the *asynchronous mode* in which the automaton that wants to move right is suspended and only the other automaton is simulated; in fact, the construction becomes simpler if we switch to the asynchronous mode, even if both automata agree on moving to the left. The position on the input where this divergence happens is called *synchronization point*. As soon as we reach the synchronization point again, we resume the suspended automaton if now both automata agree on going to the right, else we re-enter the asynchronous mode. Note that in a complete po2-Büchi automaton we must eventually re-visit the synchronization point. Now, the problem is that during the simulation of the active automaton we must be able to check "on the fly" whether the synchronization point is reached again. Therefore, instead of simulating the original automaton, we use the one from Lemma 8, which allows us to recognize when the synchronization point is reached again.

Next, we give the formal construction of the automaton $\mathcal{B}$. Let $\mathcal{A}_i = (Z_i, \Gamma, \delta_i, x_i^0, F_i)$ with $Z_i = X_i \dot\cup Y_i$ for $i \in \{1, 2\}$. We set $\mathcal{B} = (Z, \Gamma, \delta, x^0, F)$ with $Z = X \dot\cup Y$ and:

- $Z \subseteq (X_1 \times X_2 \times \Gamma^*) \cup (Z_1 \times Z_2 \times \Gamma^* \times \mathbb{N} \times \{\mathcal{A}_1, \mathcal{A}_2\})$. States from the first term of the union are those of the synchronous mode, whereas the remaining states are of the asynchronous mode. In both cases, the third component is a stack which contains words $w \in \Gamma^*$ with $|w| \le m$. For the asynchronous states the fourth component stores a counter which is bounded by the current stack size. The fifth component specifies which automaton is currently being simulated. We say that this automaton is *active*.

- $Y = ((Y_1 \times Z_2 \times \Gamma^* \times \mathbb{N} \times \{\mathcal{A}_1\}) \cup (Z_1 \times Y_2 \times \Gamma^* \times \mathbb{N} \times \{\mathcal{A}_2\})) \cap Z$ and $X = Z \setminus Y$, i.e., the left-moving states are those asynchronous states with a left-moving state of the active automaton.

- $x^0 = (x_1^0, x_2^0, \varepsilon)$, so the computation starts in the synchronous mode with an empty stack and both automata are in their respective initial states.

- For recognizing the union we set $F = ((F_1 \times X_2 \times \Gamma^*) \cup (X_1 \times F_2 \times \Gamma^*)) \cap Z$.

- For recognizing the intersection we set $F = (F_1 \times F_2 \times \Gamma^*) \cap Z$.

The transition function $\delta$ of $\mathcal{B}$ is given as follows. First, suppose that $z = (x_1, x_2, v)$ is a synchronous state and let $x_1 \xrightarrow{c} z_1$ in $\mathcal{A}_1$ and $x_2 \xrightarrow{c} z_2$ in $\mathcal{A}_2$. If a state change happens, then $c$ is pushed to the stack, i.e., we define $v' = vc$ if $z_1 \ne x_1$ or $z_2 \ne x_2$ and $v' = v$ otherwise. We set

$$(x_1, x_2, v) \xrightarrow{c} \begin{cases} (z_1, z_2, v') & \text{if } z_1 \in X_1 \text{ and } z_2 \in X_2, \\ (z_1, z_2, v', |v'|, \mathcal{A}_1) & \text{if } z_1 \in Y_1, \\ (x_1, z_2, v', |v'|, \mathcal{A}_2) & \text{else,} \end{cases}$$

that is, we stay in synchronous mode if $\mathcal{A}_1$ and $\mathcal{A}_2$ agree on moving to the right. Otherwise we make one of the left-moving automata active. If both of them want to move to the left, then $\mathcal{A}_1$ is given precedence. Now, $v'$ is a stack of labels of positions, at which at least one of the automata has changed its state. A crucial observation is the following: If $v' = a_1 \cdots a_m$ and $\alpha = u_1 a_1 \cdots u_m a_m \beta$ is the factorization induced by the positions corresponding to the $a_i$'s, then this is the $v'$-prefix factorization of $\alpha$. Moreover, since a change of direction involves a change of state, the position corresponding to $a_m$ is precisely the synchronization point for the asynchronous mode.

We now describe the transitions for an asynchronous state $(z_1, z_2, v, k, \mathcal{A}_i)$. By symmetry, we can assume $\mathcal{A}_i = \mathcal{A}_1$. Let $\mathcal{C}$ be the automaton from Lemma 8 for $\mathcal{A}_1$ and $v$. We add transitions $(z_1, z_2, v, k, \mathcal{A}_1) \xrightarrow{c} (z_1', z_2, v, k', \mathcal{A}_1)$ to another asynchronous state if $(z_1, v, k) \xrightarrow{c} (z_1', v, k')$ in $\mathcal{C}$. We therefore simulate the active automaton and moreover, by the properties of the automaton $\mathcal{C}$, we know that, in a state $(z_1, z_2, v, k, \mathcal{A}_1)$, the head position has reached the synchronization point if and only if $z_1 \in X_1$, $k = |v|$ and the current letter $c$ is $v(k)$. If this is the case and if $z_1 \xrightarrow{c} z_1'$ in $\mathcal{A}_1$ and $z_2 \xrightarrow{c} z_2'$ in $\mathcal{A}_2$, then we set

$$(z_1, z_2, v, k, \mathcal{A}_1) \xrightarrow{c} \begin{cases} (z_1', z_2', v) & \text{if } z_1' \in X_1 \text{ and } z_2' \in X_2, \\ (z_1', z_2, v, |v|, \mathcal{A}_1) & \text{if } z_1' \in Y_1, \\ (z_1, z_2', v, |v|, \mathcal{A}_2) & \text{else,} \end{cases}$$

i.e., we switch to the synchronous mode if now both automata agree on moving to the right. Otherwise, we re-enter the asynchronous mode and again one of the automata is suspended.

The constructions employed are effective and this automaton is partially ordered. Note that each time the asynchronous mode is entered or re-entered, a state change in at least one of the automata happens. There are at most $m$ state changes in synchronous mode. Thus the stack size is bounded by $m$. This yields the term $|\Gamma|^{m+1}$ in the bound for $|Z|$. □

**Corollary 10** *The class of languages recognized by deterministic po2-Büchi automata is effectively closed under Boolean operations.*

*Proof:* Deterministic po2-Büchi automata can be made complete. Therefore, effective closure under Boolean operations follows by Lemma 5 and Proposition 9. □

**Proposition 11** *Every restricted unambiguous $\omega$-monomial is recognized by a deterministic po2-Büchi automaton.*

*Proof:* Let $L = A_1^* a_1 \cdots A_k^* a_k A_{k+1}^\omega$ be an unambiguous $\omega$-monomial with $\{a_i, \ldots, a_k\} \not\subseteq A_i$ for all $1 \leq i \leq k$. This implies $a_i \notin A_1$ for some $i \geq 1$. Let $i$ be minimal with this property. For each $\alpha \in L$ we consider the $a_i$-prefix factorization $\alpha = u a_i \beta$ with $a_i \notin \mathrm{alph}(u)$. There are two cases:

$$\begin{aligned} &u \in A_1^* a_1 \cdots A_i^*, & &\beta \in A_{i+1}^* a_{i+1} \cdots A_k^* a_k A_{k+1}^\omega & &\text{or} \\ &u \in A_1^* a_1 \cdots A_j^*, & a_i \in A_j, & \beta \in A_j^* a_j \cdots A_k^* a_k A_{k+1}^\omega \end{aligned}$$

with $2 \leq j \leq i$. In each case, the expression $Q = A_j^* a_j \cdots A_k^* a_k A_{k+1}^\omega$ is unambiguous because $L$ is. Moreover, it has a degree that is smaller than that of the expression for $L$, and we have $\{a_\ell, \ldots, a_k\} \not\subseteq A_\ell$ for all $j \leq \ell \leq k$. By induction, $Q$ is recognized by some complete deterministic po2-Büchi automaton $\mathcal{B}$. The unambiguous monomial $P = A_1^* a_1 \cdots A_j^* \cap (\Gamma \setminus \{a_i\})^*$ is accepted by a deterministic po2-automaton $\mathcal{A}$ operating on finite words [15]. We modify this automaton in order to use the letter $a_i$ instead of $\triangleleft$ as a right end marker.

From these two automata $\mathcal{A}$ and $\mathcal{B}$, we construct a deterministic po2-Büchi automaton $\mathcal{C}$ accepting the $\omega$-language $PaQ$ with $a = a_i$. First, $\mathcal{C}$ checks whether there exists some $a$-position. If so, then $\mathcal{C}$ returns to the first letter of the word and starts a simulation of $\mathcal{A}$. If this automaton accepts the word, i.e., $u \in P$, then $\mathcal{C}$ moves its head to the position after the first $a$-position and starts an automaton $\widehat{\mathcal{B}}$. This automaton simulates $\mathcal{B}$ but ensures that a state change from $z$ into another state is only successfully performed if it does not happen to the left of the first $a$-position. This is done by scanning for an $a$-position to the left. Afterwards we return to the position where this test was initiated from. This is possible since $\mathcal{B}$ is deterministic. Finally we perform the transition from $z$ if the test was successful. If the test failed, then we know that we have overrun the first $a$-position in $z$. Therefore, in this case, we return to the first $a$-position, and we use the transition which $\mathcal{B}$ would have done when reading the left end marker $\triangleright$ instead of $a$. There are at most $i$ cases from above for a word $\alpha \in L$. Therefore, $L$ is a finite union of languages of the form $PaQ$ recognized by deterministic po2-Büchi automata. Proposition 9 implies the claim.

In what follows, we describe the construction of $\widehat{\mathcal{B}}$ from $\mathcal{B}$. Let $z_1 \sqsubseteq \cdots \sqsubseteq z_n$ be a linear ordering of the states $Z$ of $\mathcal{B}$ with $z_1$ being the initial state of $\mathcal{B}$. We inductively construct a sequence of deterministic po2-Büchi automata $\mathcal{B}_0, \ldots, \mathcal{B}_n$ with $\mathcal{B}_j = (Z_j, \Gamma, \delta_j, z_1, F)$ and $F$ being the set of final states of $\mathcal{B}$. The above sequence of automata will satisfy the following invariants: $Z = Z_0 \subseteq \cdots \subseteq Z_n$, $\emptyset = \delta_0 \subseteq \cdots \subseteq \delta_n$, and for all $0 \leq j \leq n$, if

$$(z_1, 1) = (q_0, i_0) \vdash_\mathcal{B} \cdots \vdash_\mathcal{B} (q_t, i_t)$$

on input $\beta \in \Gamma^\omega$ with $t \geq 1$ and $q_{t-1} \sqsubseteq z_j$, then

$$(q_0, r + i_0) \vdash_{\mathcal{B}_j}^* \cdots \vdash_{\mathcal{B}_j}^* (q_t, r + i_t)$$

on any input $ua\beta$ with $a \notin \mathrm{alph}(u)$ and $r = |ua|$. Here, $\vdash_{\mathcal{B}_j}^*$ denotes a sequence of transitions in $\mathcal{B}_j$. It follows that computations of $\mathcal{B}$ on input $\beta$ using only states up to $z_j$ are relativized in $\mathcal{B}_j$ to the suffix $\beta$ of $ua\beta$. Therefore, $\widehat{\mathcal{B}} = \mathcal{B}_n$ is the desired automaton.

Suppose that we have already constructed the automaton $\mathcal{B}_{j-1}$. We want to construct $\mathcal{B}_j$. All transitions from $\mathcal{B}_{j-1}$ are taken over and moreover, we add to $\mathcal{B}_j$ a disjoint copy of $\mathcal{B}_{j-1}$ denoted by $\mathcal{B}'_{j-1}$. We will use this copy later on. If $z_j \in X$, then we add all transitions $z_j \xrightarrow{c} z$ from $\mathcal{B}$. Let now $z_j \in Y$ and $z_j \xrightarrow{c} z$ in $\mathcal{B}$. If $z = z_j$, then we add the transition to $\mathcal{B}_j$. Suppose $z \neq z_j$. In this case we add states and transitions to $\mathcal{B}_j$ that scan for an $a$-position which is smaller than the current position. If no $a$ is found, we add to $\mathcal{B}_j$ a transition to a new $X$-state $x$ and transitions $x \xrightarrow{c} x$ for $c \neq a$ and $x \xrightarrow{a} x'$ if $z_j \xrightarrow{\triangleright} x'$ in $\mathcal{B}$. The effect is that $\mathcal{B}_j$ goes to the first $a$-position and at this position, it behaves as if $\mathcal{B}$ was reading $\triangleright$. Note that in this case the first $a$-position was last read in state $z_j$.

If an $a$ is found to the left, then we have to go back to where we started the test. For this, $\mathcal{B}_j$ uses the copy $\mathcal{B}'_{j-1}$, i.e., $\mathcal{B}_j$ goes to the position after the first $a$-position and starts $\mathcal{B}'_{j-1}$ in its initial state. Let $z'_j$ be the state of the copy corresponding to $z_j$. For all self-loops $z_j \xrightarrow{b} z_j$ in $\mathcal{B}$ we add $z'_j \xrightarrow{b} z'_j$ to $\mathcal{B}_j$, and for $z_j \xrightarrow{c} z$ in $\mathcal{B}$ we add the transition $z'_j \xrightarrow{c} z$ to $\mathcal{B}_j$. □

The following lemma shows the converse of Proposition 11. Our proof reuses techniques from the proof of Lemma 3, which in turn yields a different proof as the usual one for finite words [15].

**Lemma 12** *The language recognized by a deterministic po2-Büchi automaton is a finite union of restricted unambiguous $\omega$-monomials.*

*Proof:* Let $\mathcal{A}$ be a deterministic po2-Büchi automaton and let $\alpha \in L(\mathcal{A})$. We consider the accepting computation of $\mathcal{A}$ on $\alpha$. For this computation, we define the factorization $\alpha = u_1 a_1 \cdots u_k a_k \beta$ with $a_i \in \Gamma$, $u_i \in \Gamma^*$, and $\beta \in \Gamma^\omega$ such that the positions of the markers $a_i$ are exactly those where a state change happens in the computation. In each traversal of one of the factors $u_i$ and of the suffix $\beta$, the letters in these factors correspond to self-loops at the respective states in the accepting computation. Thus $P_\alpha = A_1^* a_1 \cdots A_k^* a_k B^\omega \subseteq L(\mathcal{A})$ for $A_i = \mathrm{alph}(u_i)$ and $B = \mathrm{alph}(\beta)$. Moreover, $P_\alpha$ is unambiguous since $\mathcal{A}$ is deterministic. Assume $\{a_i, \ldots, a_k\} \subseteq A_i$ and consider the first $X$-state $q$ of $\mathcal{A}$ entered after reading the marker $a_{i-1}$ in the above factorization of $\alpha$. There is a loop at $q$ for all letters $a_i, \ldots, a_k$. Hence, there cannot be any further markers after $a_{i-1}$. This contradicts the definition of $a_i, \ldots, a_k$. Therefore, $P_\alpha$ is a restricted unambiguous $\omega$-monomial. It follows that $L(\mathcal{A})$ is the union of all $P_\alpha$ ranging over $\alpha \in L(\mathcal{A})$. This union is finite since the degree of each $\omega$-monomial $P_\alpha$ is bounded by the number of states in $\mathcal{A}$ and there are only finitely many $\omega$-monomials of bounded degree. □

**Theorem 13** *Let $L \subseteq \Gamma^\omega$. The following assertions are equivalent:*

1. *$L$ is recognized by a deterministic po2-Büchi automaton.*
2. *$L$ is definable in $\Delta_2$.*

*Proof:* An $\omega$-language $L$ is $\Delta_2$-definable if and only if $L$ is a finite union of restricted unambiguous $\omega$-monomials [7]. The implication "1 ⇒ 2" is Lemma 12. For "2 ⇒ 1" let $L$ be a finite union of restricted unambiguous $\omega$-monomials. Proposition 11 shows that each of these $\omega$-monomials is recognized by a deterministic po2-Büchi automaton, and Proposition 9 yields an automaton for their union. □

**Corollary 14** *For a given regular $\omega$-language $L$ it is decidable whether $L$ can be recognized by some deterministic po2-Büchi automaton.*

*Proof:* By Theorem 13, $L$ is recognizable by some deterministic po2-Büchi automaton if and only if $L$ is $\Delta_2$-definable. The latter is known to be decidable [2, 7]. □

**Example 15** The $\omega$-language $L = \{a,b\}^* a \emptyset^* c \{c\}^\omega$ is a restricted unambiguous $\omega$-monomial and thus definable in $\Delta_2$, see [7]. By Theorem 13 it is recognized by a deterministic po2-Büchi automaton. Moreover, $L$ cannot be recognized by a deterministic partially ordered *one-way* Büchi automaton. Hence, the class of $\omega$-languages recognizable by deterministic partially ordered one-way Büchi automata is a strict subclass of the class recognizable by deterministic po2-Büchi automata. ◇

## 5 Complexity Results

In this section, we show that several decision problems for po2-Büchi automata are coNP-complete. This is surprising since for general Büchi automata, these problems are PSPACE-hard [16]. We consider the following decision problems for given po2-Büchi automata $\mathcal{A}$ and $\mathcal{B}$:

- INCLUSION: Decide whether $L(\mathcal{A}) \subseteq L(\mathcal{B})$.

- EQUIVALENCE: Decide whether $L(\mathcal{A}) = L(\mathcal{B})$.

- EMPTINESS: Decide whether $L(\mathcal{A}) = \emptyset$.

- UNIVERSALITY: Decide whether $L(\mathcal{A}) = \Gamma^\omega$.

**Theorem 16** EMPTINESS *is* coNP*-complete for both nondeterministic and deterministic po2-Büchi automata.* INCLUSION, EQUIVALENCE *and* UNIVERSALITY *are* coNP*-complete for deterministic po2-Büchi automata; for* INCLUSION *it suffices that* $\mathcal{B}$ *is deterministic.*

The proof of this theorem can be found at the end of this section.

**Lemma 17** INCLUSION *is in* coNP *for nondeterministic* $\mathcal{A}$ *and deterministic* $\mathcal{B}$.

*Proof:* Let $Z_\mathcal{A}$ and $Z_\mathcal{B}$ be the states of $\mathcal{A}$ and $\mathcal{B}$, respectively. We have $L(\mathcal{A}) \subseteq L(\mathcal{B})$ if and only if $L(\mathcal{A}) \setminus L(\mathcal{B}) = \emptyset$. By Lemma 5 we see that we can easily compute a deterministic po2-Büchi automaton $\overline{\mathcal{B}}$ such that $L(\overline{\mathcal{B}}) = \Gamma^\omega \setminus L(\mathcal{B})$. If $L(\mathcal{A}) \cap L(\overline{\mathcal{B}}) \neq \emptyset$, then, by Lemma 3, there is a word $u$ with $|u| \leq |Z_\mathcal{A}| + |Z_\mathcal{B}|$ and a letter $a \in \Gamma$ such that $ua^\omega \in L(\mathcal{A}) \cap L(\overline{\mathcal{B}}) = L(\mathcal{A}) \setminus L(\mathcal{B})$. We might have to add one state in each of $\mathcal{A}$ and $\mathcal{B}$ for making them complete. Therefore, in order to test $L(\mathcal{A}) \not\subseteq L(\mathcal{B})$, it suffices to guess a word $u$ of length at most $|Z_\mathcal{A}| + |Z_\mathcal{B}|$ and a letter $a \in \Gamma$ with $ua^\omega \in L(\mathcal{A}) \cap L(\overline{\mathcal{B}})$. Hence, non-inclusion can be verified in NP, i.e., INCLUSION is in coNP. □

**Lemma 18** EMPTINESS *is* coNP*-hard for deterministic po2-Büchi automata.*

*Proof:* We shall reduce the complement of SAT to EMPTINESS. Let $\varphi$ be a propositional formula and let $v_1, \ldots, v_m$ be the variables used in $\varphi$. We give the construction of a deterministic po2-automaton $\mathcal{A}_\varphi$ over the alphabet $\{0, 1\}$ such that $L(\mathcal{A}_\varphi) = \emptyset$ if and only if there is no satisfying assignment for $\varphi$. The idea is that we identify the position $i$ of the input with the assignment of variable $i$ for $1 \leq i \leq m$. The rest of the input has no effect on the computation.

Inductively we construct an automaton with the following characteristics: There are two distinguished X-states $x_t$ and $x_f$ with a loop for both letters 0 and 1. No other right-moving state has a self-loop. The state $x_t$ is eventually entered if $\varphi$ evaluates to true under the input, else $x_f$ is eventually entered. Moreover, $x_t$ and $x_f$ are only entered by transitions reading $\triangleright$ and $x_t$ is the sole final state. Hence an input is accepted if and only if eventually $x_t$ is entered. In case it is rejected, it eventually enters $x_f$.

For variables $v_i$ the automaton $\mathcal{A}_{v_i}$ skips the first $i-1$ letters of the input, remembers the letter $a_i$ at position $i$ and returns to the beginning of the word. If $a_i = 1$ then $\mathcal{A}$ enters $x_t$ else it enters $x_f$. For the negation we simply swap the states $x_f$ and $x_t$. For $\varphi \wedge \psi$ we compose the automata $\mathcal{A}_\varphi$ and $\mathcal{A}_\psi$ in the following way: The states $x_t$ and $x_f$ of $\mathcal{A}_\varphi$ are deleted. Transitions of $\mathcal{A}_\varphi$ leading into state $x_f$ are redirected to the corresponding state $x_f$ of $\mathcal{A}_\psi$; transitions leading into state $x_t$ are redirected to the initial state of $\mathcal{A}_\psi$. Similarly, we get an automaton for $\varphi \vee \psi$. □

*Proof (Proof of Theorem 16):* For nondeterministic $\mathcal{A}$ and deterministic $\mathcal{B}$, Lemma 17 shows that INCLUSION is in coNP. Therefore, EMPTINESS, EQUIVALENCE and UNIVERSALITY for deterministic po2-Büchi automata are in coNP, too. Lemma 18 yields coNP-hardness of EMPTINESS for deterministic po2-Büchi automata. Closure under complement (Lemma 5) yields a reduction from EMPTINESS to UNIVERSALITY. Together with the straightforward reductions from UNIVERSALITY to EQUIVALENCE and from EMPTINESS to INCLUSION, this yields coNP-hardness of all these problems for deterministic po2-Büchi automata. □

## Acknowledgments

This work was supported by the German Research Foundation (DFG) under grant DI 435/5-1.